\documentclass[%
preprint,
 amsmath,amssymb,
 aps,
pre,
citeautoscript,
]{revtex4-2}
\usepackage{graphicx}
\usepackage{dcolumn}
\usepackage{bm}

\begin{document}


\title{Machine Learning of consistent thermodynamic models using automatic differentiation.}

\affiliation{Los Alamos National Laboratory, Theoretical Division, Physics and Chemistry of Materials Group, Los Alamos, 87545 New Mexico (USA). }
\affiliation{ 
Los Alamos National Laboratory, Computer, Computational \& Stat Sciences Division,  Information Sciences Group, Los Alamos, 87545 New Mexico (USA).}
\author{David Rosenberger}
 \email{d.rosenberger@fu-berlin.de}
 \altaffiliation[Current affiliation:~]{Freie Universit\"at Berlin (Germany).}
 \affiliation{Los Alamos National Laboratory, Theoretical Division, Physics and Chemistry of Materials Group, Los Alamos, 87545 New Mexico (USA). }
\author{Kipton Barros}
\affiliation{Los Alamos National Laboratory, Theoretical Division, Physics and Chemistry of Materials Group, Los Alamos, 87545 New Mexico (USA). }
\author{Timothy C. Germann}
\affiliation{Los Alamos National Laboratory, Theoretical Division, Physics and Chemistry of Materials Group, Los Alamos, 87545 New Mexico (USA). }

\author{Nicholas Lubbers}%
\email{nlubbers@lanl.gov}
\affiliation{ 
Los Alamos National Laboratory, Computer, Computational \& Statistical Sciences Division,  Information Sciences Group, Los Alamos, 87545 New Mexico (USA).}

\begin{abstract}
We propose a data-driven method to describe consistent equations of state (EOS) for arbitrary systems. Complex EOS are traditionally obtained by fitting suitable analytical expressions to thermophysical data. A key aspect of EOS are that the relationships between state variables are given by derivatives of the system free energy. In this work, we model the free energy with an artificial neural network, and utilize automatic differentiation to directly learn the derivatives of the free energy. We demonstrate this approach on two different systems, the analytic van der Waals EOS, and published data for the Lennard-Jones fluid, and show that it is advantageous over direct learning of thermodynamic properties (i.e. not as derivatives of the free energy, but as independent properties), in terms of both accuracy and the exact preservation of the Maxwell relations. Furthermore, the method implicitly provides the free energy of a system without explicit integration.
\end{abstract}

\maketitle

\section{Introduction}
Thermophysical properties of liquids
are routinely measured in laboratory and computational experiments and then summarized in corresponding data sets~\cite{gammon_dippr_1997,linstrom_nist_1997,marrero_solubility_2003,royal_chemical_society_chemspider_nodate}. However, the amount of available data is small compared to the total number of systems of interest for both, pure components and mixtures~\cite{ruddigkeit_enumeration_2012}. Given the time and resource cost of direct measurements, scientists have used empirical functional forms for Equations of State (EOS), which model the relationship of state variables like density, pressure and temperature~\cite{muller_history_2007,rowlinson_cohesion_2005,ungerer_molecular_2007}. The derivation of those functional forms has been driven by a combination of physical laws, statistical mechanics, and choice of empirical functional forms~\cite{kontogeorgis_thermodynamic_2010,nieto-draghi_general_2015}. Although these approaches have proven useful, limitations in applicability of statistical mechanics for complex systems and in choice of functional form may hinder the development of a general ansatz to estimate EOS.  
Therefore, we propose to utilize automatic differentiation (AD)~\cite{griewank_a_automatic_1989,baydin_atilim_gune_s_automatic_2018} in combination with artificial neural networks (ANNs)~\cite{lecun_deep_2015} to model thermophysical properties and to implicitly compute the free energy of a system. AD consists of a set of high-level algorithms for efficiently computing certain derivatives of computations without numerical approximation. The goal of the proposed method is to learn the properties of interest as a derivative of a model free energy with respect to density, pressure, and temperature, thereby ensuring the thermodynamic consistency of the resulting model. This process can be applied regardless of the availability of free energy data in the examined dataset.

A similar philosophy has been utilized by Christensen et al. to learn potential energy surfaces for atoms using response operators~\cite{christensen_operators_2019}. 
Rather than kernel-based machine learning (ML) methods applied by Christensen et al., we employ neural networks, which is advantageous as the response functions are automatically generated.

ML approaches to fitting EOS have recently been studied~\cite{zhu_generating_2020,dai_efficient_2020}. EOS fits are often achieved by direct minimization of the difference between a measured property and its corresponding predicted value~\cite{liu_machine_2019,craven_machine_2020}. Although straightforward, this direct approach can lead to inconsistencies in the EOS via violation of the Maxwell relations, as we will show. Another approach to model EOS is to train ML models to the function that, when differentiated, generates the quantities of interest. Those functions can either be particle-particle interactions~\cite{berressem_boltzmann_2021}, or the partition function~\cite{desgranges_ensemble_2020}, or the system's free energy~\cite{wu_solving_2019}. 
However, those methods rely on accurate sampling of the phase space, which for most systems is a non-trivial task. Therefore, we propose an indirect method to model the free energy of a system from data on the free energy derivatives, which are easier to compute and thus effectively reduce the number of expensive and complex computations of the phase space. The power of neural networks to estimate a free energy has been demonstrated by Nicoli et al. in the context of lattice fluid theory~\cite{PhysRevLett.126.032001}. The authors train an estimator in order to predict the free energy from data sets of microscopic configurations. This is conceptually different to the method proposed here, which learns from thermodynamic observables that can be, for example, measured in experiment. We suggest to train a NN in order to predict derivatives of the free energy and by doing so implicitly matching the underlying free energy.
\section{Method}
While the model could be based in any thermodynamic potential, in this work we choose to start with the Helmholtz free energy $A$ defined in the canonical ensemble: 
\begin{equation}
\label{free_ener}
    A(N,V,T) = U-TS.
\end{equation}
$A$ is a function of the particle number $N$, the volume $V$, and temperature $T$. $U$ is the internal energy of a system and $S$ is the entropy. 
Thermophysical properties such as pressure $P$, chemical potential $\mu$ and the entropy $S$ can be computed from 1st order partial derivatives  of the free energy:
$
P= - \left(\frac{\partial A}{\partial V}\right)_{N,T}\label{P_from_A}$,
$\mu= \left(\frac{\partial A}{\partial N}\right)_{V,T}$,
$S=  - \left(\frac{\partial A}{\partial T}\right)_{V,T}$,
$U = A +TS=A-T \left(\frac{\partial A}{\partial T}\right)_{V,T}$.
Further, properties like the isothermal compressibility $\beta$, the thermal pressure coefficient $\gamma$, the isochoric heat capacity $c_v$, and the thermal expansion coefficient $\alpha$ can be computed from first order derivatives of the thermophysical properties $P$ and $U$.
\begin{eqnarray}
\beta&=& -\frac{1}{V} \left(\frac{\partial V}{\partial P}\right)_{N,T} =  \frac{1}{V} \left(\frac{\partial \left(\frac{\partial A}{\partial V}\right)_{N,T}}{\partial V}\right)_{N,T}^{-1} \label{iso_compress}\\
\gamma&=&\left(\frac{\partial P}{\partial T}\right)_{V,T}=\left(\frac{-\partial \left(\frac{\partial A}{\partial V}\right)_{N,T}} {\partial T}\right)_{V,T}\label{thermal_press}\\
c_v&=&\left(\frac{\partial U}{\partial T}\right)_{V,T}=\left(\frac{\partial \left(A-T\left(\frac{\partial A}{\partial T}\right)_{V,T}\right)}{\partial T}\right)_{V,T}\label{isochor_heat}\\
\alpha&=&\beta\cdot\gamma
\end{eqnarray}

While these relations are well known, it is difficult to solve them in practice as either the functional form of the free energy is not known or it is too complicated to directly compute the free energy in order to apply numerical differentiation. Therefore, we propose that the free energy of a system $\hat A$ be described with a learned function $\hat{f}$ which depends on the density $\rho=N/V$, and the temperature $T$. $\hat{f}$ is expressed in form of an ANN, whose parameters $\Theta$ (weights and biases for each layer) are optimized during learning:
\begin{equation}
    \hat{A}=N \hat{f}_{\Theta}[\rho,T]
\end{equation}
The factor $N$ ensures that the learned free energy $\hat{A}$ is an extensive function, whereas the learning itself is performed on intensive variables. 

From now on we refer to this model as the Free Energy Neural Network (FE-NN) model.
The parameters $\Theta$ are optimized by minimization of the following loss function:
\begin{equation}
\begin{split}
L[\Theta]=&\,\lambda_P\left[- \left(\frac{\partial \hat{A}}{\partial V}\right)_{N,T}-P\right]^2
\\&+\lambda_{\mu}\left[\left(\frac{\partial \hat{A}}{\partial N}\right)_{V,T}-\mu\right]^2
\\&+\lambda_U\left[\hat{A}-T\left(\frac{\partial \hat{A}}{\partial T}\right)_{V,T}-U\right]^2,
\end{split}
\end{equation}
where $\hat{A}$ is the learned free energy, $\lambda_P$, $\lambda_\mu$, and $\lambda_U$ are weighting parameters whose exact forms are: 
\begin{eqnarray}
\lambda_P=\frac{1}{\frac{1}{l}\sum_{i=1}^{l} (P_i-\langle P\rangle)^2}=\frac{1}{\sigma_P} \\
\lambda_{\mu}=\frac{1}{\frac{1}{l} \sum_{i=1}^{l} (\mu_n-\langle \mu\rangle)^2}=\frac{1}{\sigma_{\mu}} \\
\lambda_U=\frac{1}{N^2\sum_{i=1}^{l} \left(\frac{U_i-\langle U\rangle}{N_i}\right)^2}=\frac{1}{N^2\sigma_{\frac{U}{N}}},
\end{eqnarray}
where $l$ equals the number of points in a corresponding data set and all variables have the same meaning as previously defined. Each term of the loss function is normalized with the corresponding variance $\sigma$. We desire to ensure that all terms in the loss are intensive, therefore for the internal energy we have to take the particle number $N$ into account. 
The loss function summarizes the main idea of this work: We are utilizing AD \emph{within the model} to seek the best approximation for the free energy $A$ subject to the known values of thermophysical properties $P$, $\mu$ and $U$. This is in addition to the standard use of AD in computing the derivative of the loss function with respect to the parameters, $\nabla_\Theta L$, to update the network; multiple, nested calls to AD are made to achieve fitting to the derivatives of the free energy.
Thus if we learn the derivatives of the free energy, we implicitly also model the free energy of the system without performing an explicit integration. At the same time, we can further utilize AD to compute second derivatives $\beta$, $\gamma$, and $c_v$. This enables a complete and consistent thermodynamic modeling technique for an arbitrary material from data.
We contrast this with a typical ML approach, where differences between target and predicted properties are directly minimized.
For this Multi-Task Neural Network (MT-NN) model the loss function takes the following form:
$
L[\Theta]=\lambda_P\left[\hat{P}-P\right]^2+\lambda_{\mu}\left[\hat{\mu}-\mu\right]^2+\lambda_U\left[\hat{U}-U\right]^2,
$
where $\hat{P}$, $\hat{U}$, and $\hat{\mu}$ indicate the properties learned by the network. The MT-NN uses the same weight functions $\lambda$ and intensive learning formulation as the FE-NN; inputs are intensive, and the extensive energy is predicted with a scaling factor of $N$.\newline
\section{Training procedure and data preparation}

\subsection{Van der Waals Equation of State}
The first test case for both ML models is the the Van der Waals EOS defined by its underlying free energy:
\begin{equation}
\label{vdW}
A (N,V,T)= -N k_B T  \left[1+\ln\left(\frac{(V-Nb)T^{3/2}}{N\Phi}\right)-\frac{aN^2}{V}\right]
\end{equation}
For constants, $a=0.01$, $b=0.01$ and $\Phi = 1.0$, we generate
1000 random state points for each independent variable.
For $N$, 1000 values are generated in the range between 0.5 and 2.
For $V$, 1000 random values are generated in the range between $3\cdot2.8$ and $3/2.8$. This leads to reduced volumes $v=V/N$ in the range between $20~v_c$ to $500~v_c$, where $v_c$ is the critical volume. Finally, for $T$ 1000 values are randomly generated in the range between $(8/27)/3$ and $(8/27)\cdot3$, leading to temperatures above and below the critical temperature of the van der Waals gas. This represents a physical but simple thermodynamic system of a low density gas.
The true values for the free energy are computed according to Eq.~\eqref{vdW} for this dataset. By applying automatic differentiation on that free energy the corresponding true values for pressure $P$, chemical potential $\mu$, internal energy $U$ and entropy $S$ are obtained. As written above, learning of the FE-NN model is performed on a per-atom (intensive) basis. We feed $\rho=N/V$ and $T$ to the neural network and all extensive target properties are normalized with the corresponding particle number.
The initial weights of the ANN are assigned with the Xavier function~\cite{pmlr-v9-glorot10a} with a gain of 1.5. Training is performed for 10000 epochs with a batch-size of 20. Each neuron is activated with the Softmax function~\cite{luce_individual_2005,bridle_john_s_training_1989}. The Adam optimizer is applied to minimize the loss function with an initial learning rate of 0.001. If the loss function does not change over 500 epochs, the learning rate is automatically decreased by a factor of 0.5.
We train both, the FE-NN  model and MT-NN model, to 20\% of the 3000 data points, use 10\% for validation, and test on the remaining 70\%. 
\subsection{Lennard-Jones system}
The data in ~\cite{stephan_thermophysical_2019} provides a suitable test case for the two ML approaches to estimate EOS. Thermodynamic properties have been collected from finite-size molecular dynamics simulation of the Lennard-Jones fluid that covers homogeneous phases, vapor-liquid equilibrium, and regions above and below the critical point of the models. However, not all data sets summarized therein contain all thermophysical properties. Therefore, we first extract all data sets which contain the following thermophysical properties: $\mu$, $P$, $U$, $\alpha$, $\beta$, $\gamma$ and $c_v$. Second, some regions in phase space are sparsely sampled and are thus difficult to learn. As such, we only include points with temperature $T<2$ and a density $\rho > 0.05$. All values are given in Lennard-Jones units. In the end, we extract data for 1721 state points.  In the data set only the excess part of properties like the chemical potential are provided, thus we manually added the ideal gas contribution where needed. Further, the data set contains only information on the density and not on particle numbers and volume. We therefore generated random particle numbers in the range between 0 and 1, and computed the volume according to $V=N/\rho$.
Training is performed in total on 70\% on the data total: 40\% are used for direct training, and 30\% are considered for validation. The test set contains the remaining 30\%. The hyperparameters for learning are the same as for the van der Waals EOS.
\section{Results}
In the following, we show results from the subsequent network architecture for the free energy model: 1 input layer with 2 neurons, 3 hidden layers with 16 neurons, and 1 output layer with 1 neuron for free energy.
The ANN for the multitask learner has the same internal structure, but one output layer each for $\hat{P}$, $\hat{U}$, and $\hat{\mu}$. Results for other architectures are presented in corresponding tables S~I-III and S~V-VII in the supporting information~\cite{supporting_information}. Our models are implemented in PyTorch~\cite{paszke_pytorch_2019}.
\subsection{Van der Waals Equations of State}
\begin{figure} [htbp]
{%
\includegraphics[width=0.4\columnwidth]{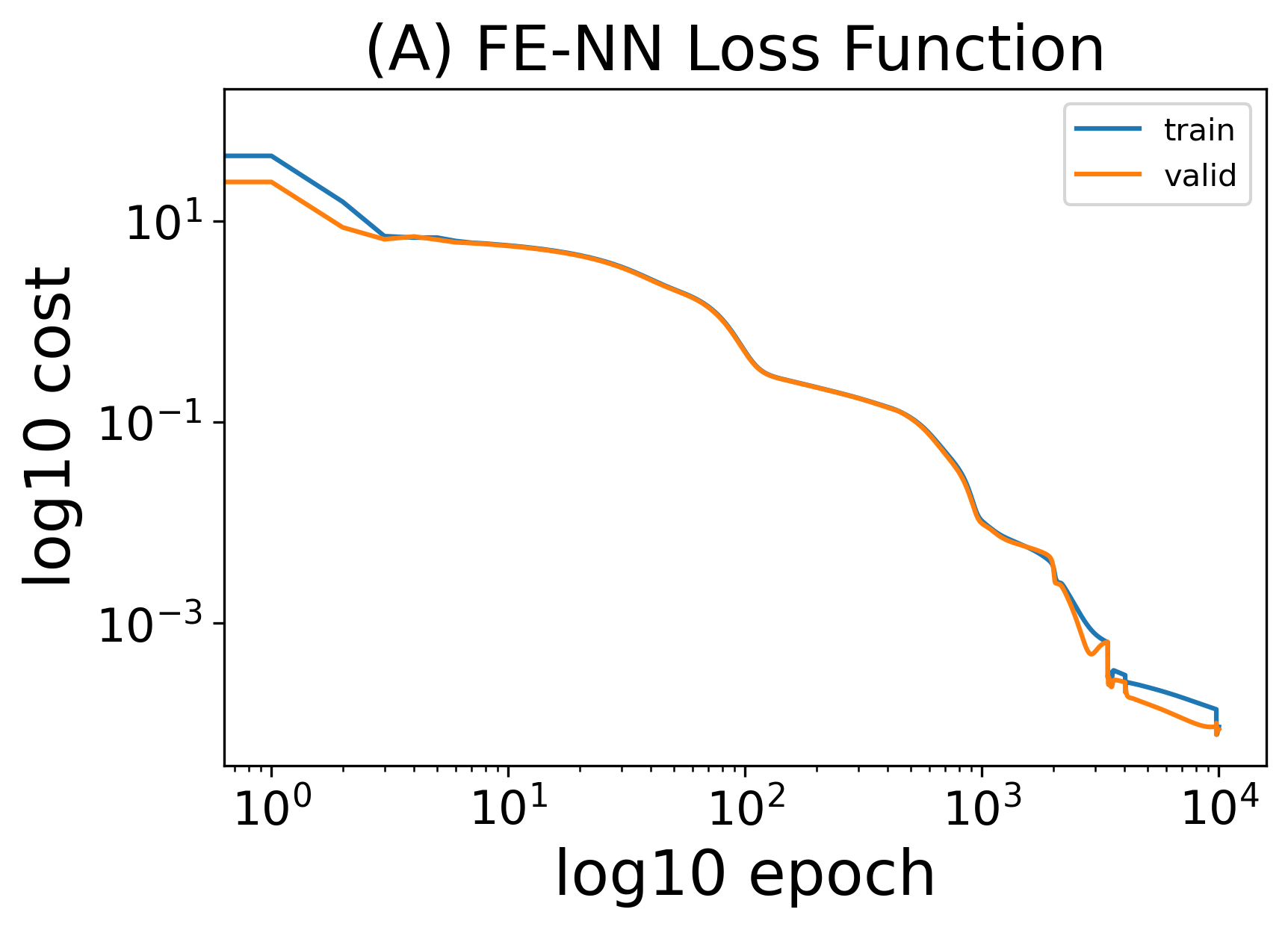}
\includegraphics[width=0.4\columnwidth]{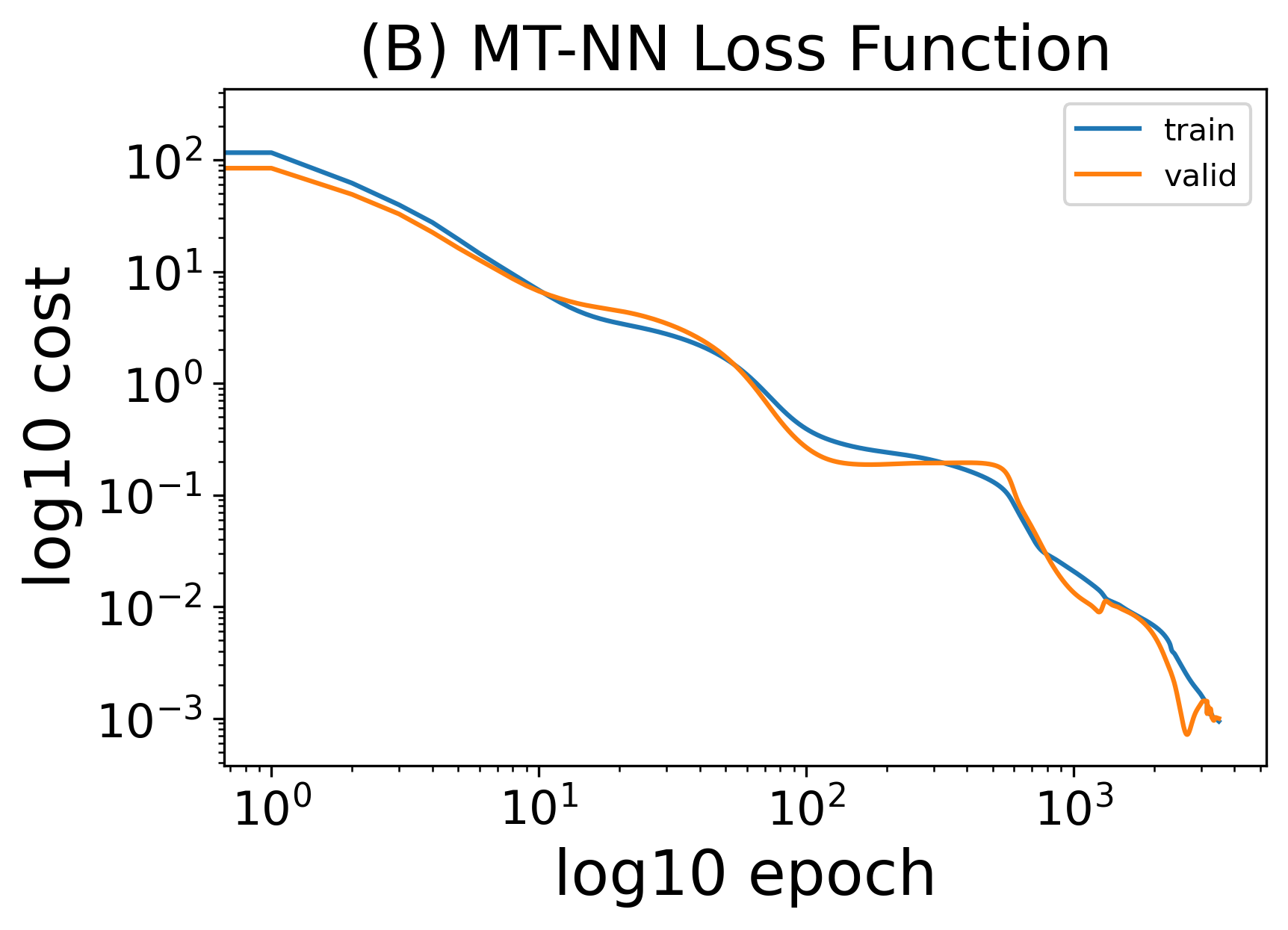}
}
\caption{(A): Loss function for the FE-NN model. Blue corresponds to the training set and orange to the validation set. (B): Loss function for the MT-NN model. Blue corresponds to the training set and orange to the validation set.}
\label{fig:cost_free_energy_vdW}
\end{figure}

We first assess the two models on the the Van der Waals (VdW) EOS. This simple test case allows a comparison of the ML models with an analytical expression. 
In Fig. \ref{fig:cost_free_energy_vdW} (A) we show the loss function of the FE-NN model as a function of the number of epochs. The blue line corresponds to the randomly selected 20\% training data, whereas the orange line corresponds to the 10\% of the training data which are used for validation. 
As one sees, the learning rate is lowered which means a low-cost combination of parameters is found.
Additionally, the final model chosen is the one that minimizes the validation error, also known as an early stopping procedure. In Fig. \ref{fig:cost_free_energy_vdW} (B) we show the loss function of the MT-NN model, which shows alike behavior.
In Table \ref{tbl:1} we present the mean absolute error (MAE) values between the ground truth and the ML predictions. The MAE values are averaged over 4 ML runs, each initialized with a different random seed for the train/validation/test splitting. For more details see Table S~IV in the supporting information~\cite{supporting_information}.
One sees that the FE-NN is consistently more accurate than the MT-NN. 
In addition we also present results for a kernel ridge regression (KRR) ML model for $\mu, P$ and $U$. This model has been established in ~\cite{craven_machine_2020} to learn $P$ and $U$ for LJ fluids. The model utilizes a polynomial kernel of the 4th order:
\begin{equation}
    K(x_i,x_j)=(\gamma x_i^T x_j+c_0)^4,
\end{equation}
where $x_i$ and $x_j$ are input features.
$\gamma$ and $c_0$ are optimized using a grid search where the optimization scoring function was the mean absolute
percent error of the corresponding thermodynamic property. The conditioning factor $\alpha$ (see Eq.~\eqref{kernel_func}) in the KRR process is set to one. As one sees, with KRR one obtains MAE values which are intermediate to the NN models for the three target properties $\mu, P$ and $U$. 

\begin{figure} [htbp]
{%
\includegraphics[width=0.8\columnwidth]{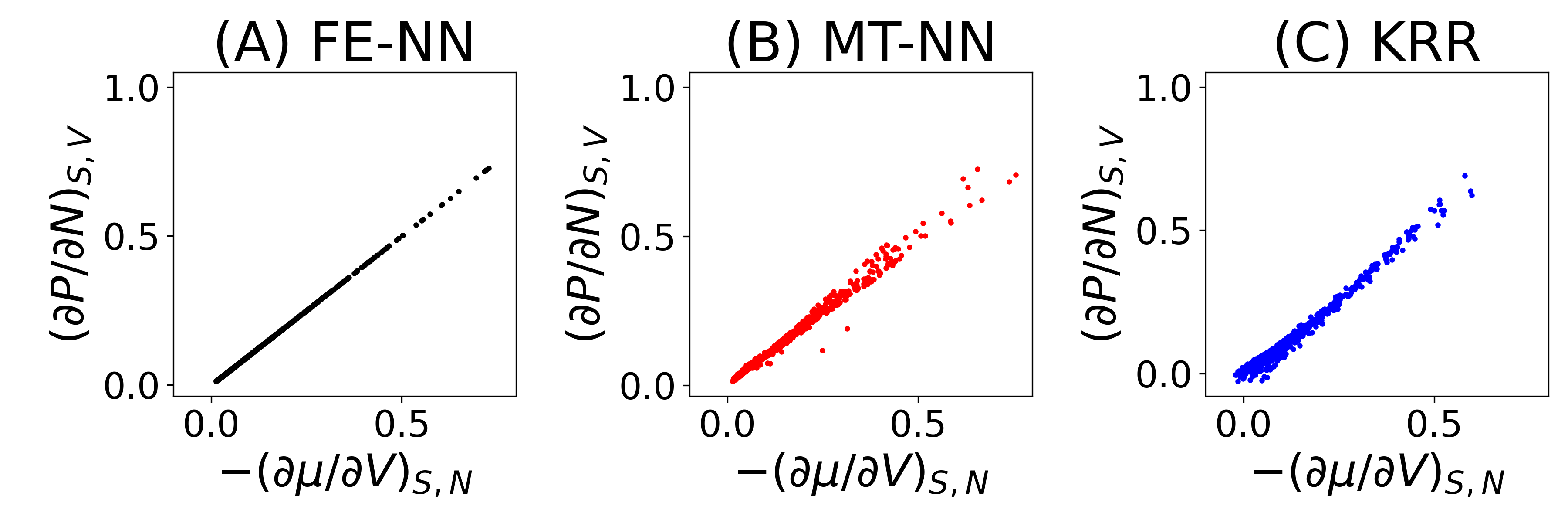}
}
\caption{Maxwell relation for the VdW EOS evaluated from (A) the FE-NN, (B) the MT-NN and (C) the KRR model.}
\label{fig:vdw_Maxwell}
\end{figure}

\begin{table}
  \caption{Mean absolute error between predicted and true data generated from the VdW EOS for the following: the chemical potential $\mu$, the pressure $P$, the internal energy $U$ and the isothermal compressibility $\beta$. For the FE-NN model, we also give the corresponding errors for the free energy $A$ and entropy $S$.}
  \label{tbl:1}
  \begin{tabular}{lllllll}
    \hline
    Model   & $\mu$  & $P$  & $U$ & $A$ & $S$ & $\beta$  \\
    \hline
    FE-NN&  0.0009 &  0.0003  & 0.0010 & 0.0005 & 0.0032 &  0.2331 \\
    MT-NN  &  0.0172 &  0.0023 & 0.0042 & \,- & \,- & 1.0270  \\
    KRR & 0.0038 & 0.0077 & 0.0002 & \,- & \,- & 16.2914\\
    \hline
  \end{tabular}
\end{table}
In Table \ref{tbl:1b}, we show the standard deviation for the MAE values computed over the different random seeds used to split the data (see Table S IV in the supplemental material for more information~\cite{supporting_information}). We see that on average the standard deviations for the FE-NN model are at least one order of magnitude smaller than for the MT-NN model. The KRR model is again intermediate between the two NN models. This confirms the higher accuracy of the FE-NN model over the MT-NN model indicated from the MAE values.
Moreover, in Table~\ref{tbl:1} we present the MAE values for the free energy $A$ and the entropy $S$ using the FE-NN. These quantities are completely unavailable in the MT-NN and KRR model, but easy to compute with the FE-NN using AD. Although $A$ and $S$ are not explicitly part of the training, the FE-NN achieves high accuracy for both properties, whereas the MT-NN and the KRR have no notion of these two quantities. 
Information regarding the free energy and entropy enter the training data only through their derivatives, and as a consequence MT-NN and KRR fail to infer prediction functions which can be consistently thermodynamically integrated, because the Maxwell relations are violated as we will show. Only the FE-NN model is able to implicitly learn the free energy of a system as it treats $\mu$, $P$ and $U$ as true derivatives of a single state function, and thus produces predictions which can be consistently thermodynamically integrated.
\begin{table}
  \caption{Standard deviations ($\sigma$) for the mean absolute error values reported in Table~\ref{tbl:1}.}
  \label{tbl:1b}
  \begin{tabular}{lllllll}
    \hline
    Model   &  $\sigma_\mu$ & $\sigma_P$  & $\sigma_U$  & $\sigma_A$  & $\sigma_S$  & $\sigma_\beta$\\
    \hline
    FE-NN & 3.37E-04 & 5.8E-05 &  4.11E-04 & 3.79E-04 & 1.871E-03 & 8.38474E-02  \\
    MT-NN  &  1.83860E-02 & 2.5586E-03 & 5.2906E-03 & \,\,- & \,\,- & 1.100733 \\
    KRR & 9.11E-04 & 1.3077E-03 & 0 & \,\,- & \,\,- & 9.84037\\
    \hline
  \end{tabular}
\end{table}
In addition, we assess the performance of the three models in predicting the isothermal compressibility. The analytical value for $\beta$ for the VdW EOS is:
\begin{equation}
\beta=-\frac{1}{V} \frac{1}{\left(\frac{\partial P}{\partial V}\right)_T}= -\frac{1}{V} \frac{1}{\frac{2aN^2}{V^3}-\frac{NkT}{(V-Nb)}}
\end{equation}
The values for the ML models are obtained from differentiation of the learned pressure with respect to the volume (see Eq.~\eqref{iso_compress}). For the NN models this is easily be obtained via AD from the network architecture, whereas for KRR we differentiate via the equation
\begin{eqnarray}
\label{kernel_func}
y_{pred}(x) = K(x,x_i) \cdot \alpha = (\gamma x_i^T x+c_0)^4 \cdot \alpha\\
\label{kernel_diff}
\frac{\partial y_{pred}(x)}{\partial q} = 4 \gamma x_j (\gamma x^T x_j +c_0)^3 \cdot \alpha \frac{\partial x}{\partial{q}},
\end{eqnarray}
where $y_{pred}$ is a predicted quantity ($\mu$, $P$, or $U$) from the Kernel model, $q$ is a thermodynamic input ($N$, $V$, or $T$), and all other variables are as previously defined.
From the MAE of $\beta$ (see Table \ref{tbl:1}) one sees that results for the FE-NN are more accurate than for the MT-NN for which the MAE is about 5 times larger. The KRR model has the largest MAE value for $\beta$.
Besides the higher accuracy in reproducing thermophysical properties, only the FE-NN is able to construct a consistent thermodynamic model.
To show this, we validate both models against the Maxwell relation:
\begin{equation}
\label{Maxwell_rel}
    \left(\frac{\partial P}{\partial N}\right)_{S,V}=-\left(\frac{\partial \mu}{\partial V}\right)_{S,N}.
\end{equation}
The derivatives for the NN model are again obtained via AD and for KRR via Eq.~\eqref{kernel_diff}.
From Fig.~\ref{fig:vdw_Maxwell}, one sees that only the FE-NN can exactly recover the Maxwell relation, whereas the other two models show deviations from exact correlation. 
This demonstrates explicitly that there is no free energy state function that corresponds to the predictions of the MT-NN and the KRR approach, highlighting the key advantage of the FE-NN approach. In light of the fact that the this dilute VdW system is so simple, this shows that there is a fundamental inconsistency that arises when applying ML to thermodynamic data without respecting the Maxwell relations, which can be overcome with the FE-NN. \newline
\subsection{Lennard-Jones Equations of State}
\begin{figure} [htbp]
{%
\includegraphics[width=0.4\columnwidth]{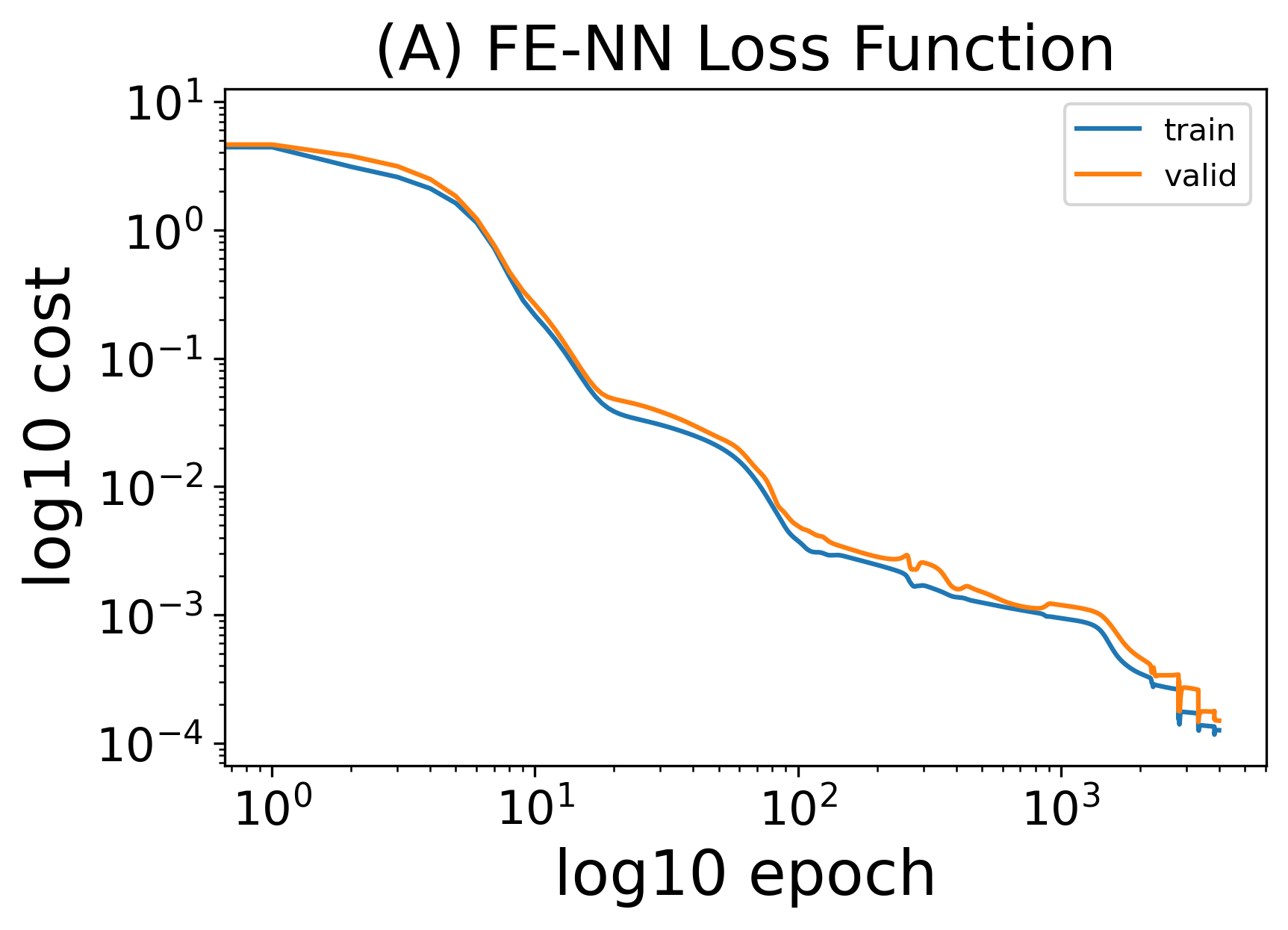}
\includegraphics[width=0.4\columnwidth]{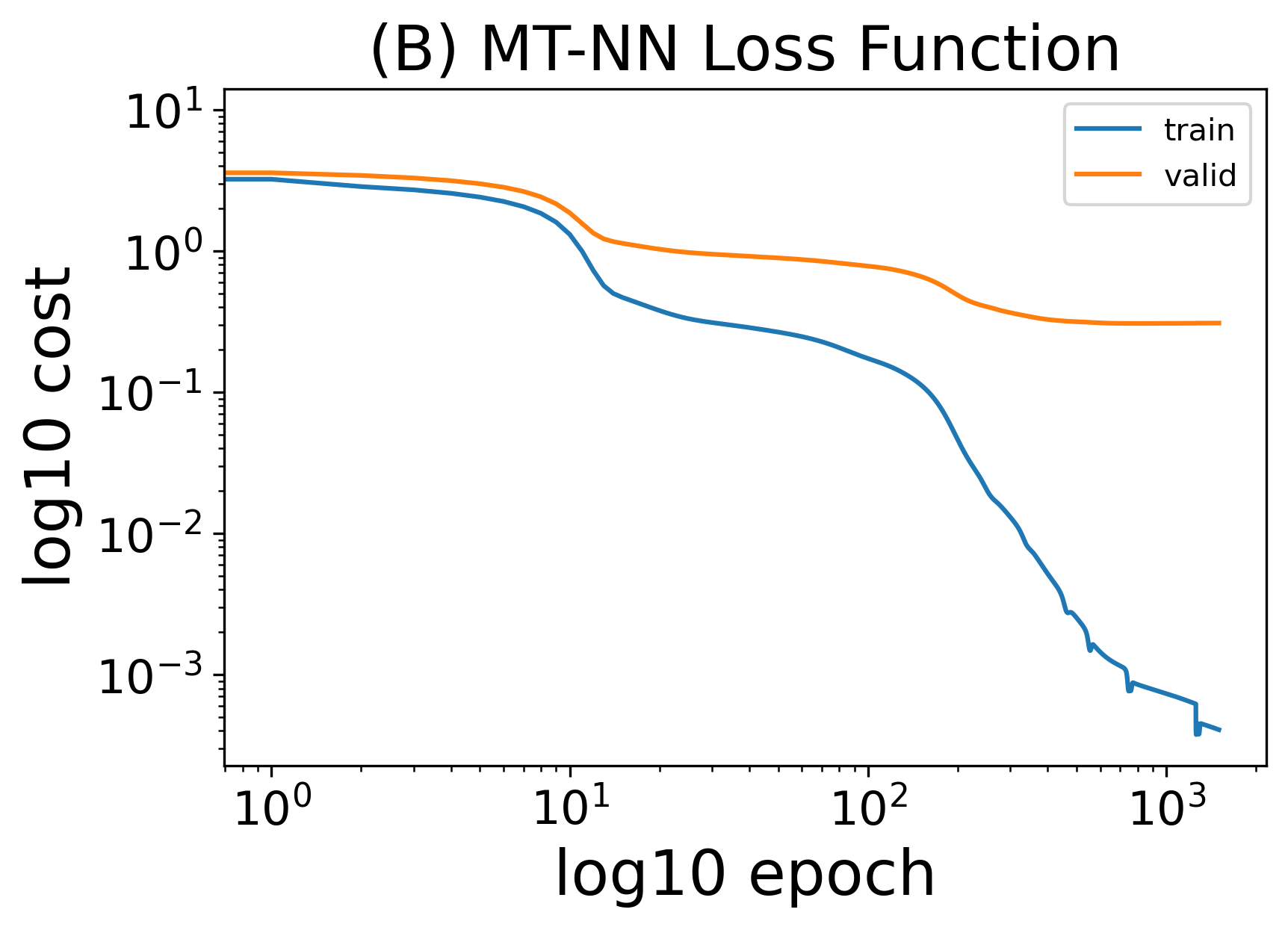}
}
\caption{(A): Loss function for the FE-NN model. Blue corresponds to the training set and orange to the validation set. (B): Loss function for the MT-NN model. Blue corresponds to the training set and orange to the validation set.}
\label{fig:LJ_free_ener_cost}
\end{figure}

\begin{figure*} [htbp]
{%
\includegraphics[width=0.8\textwidth]{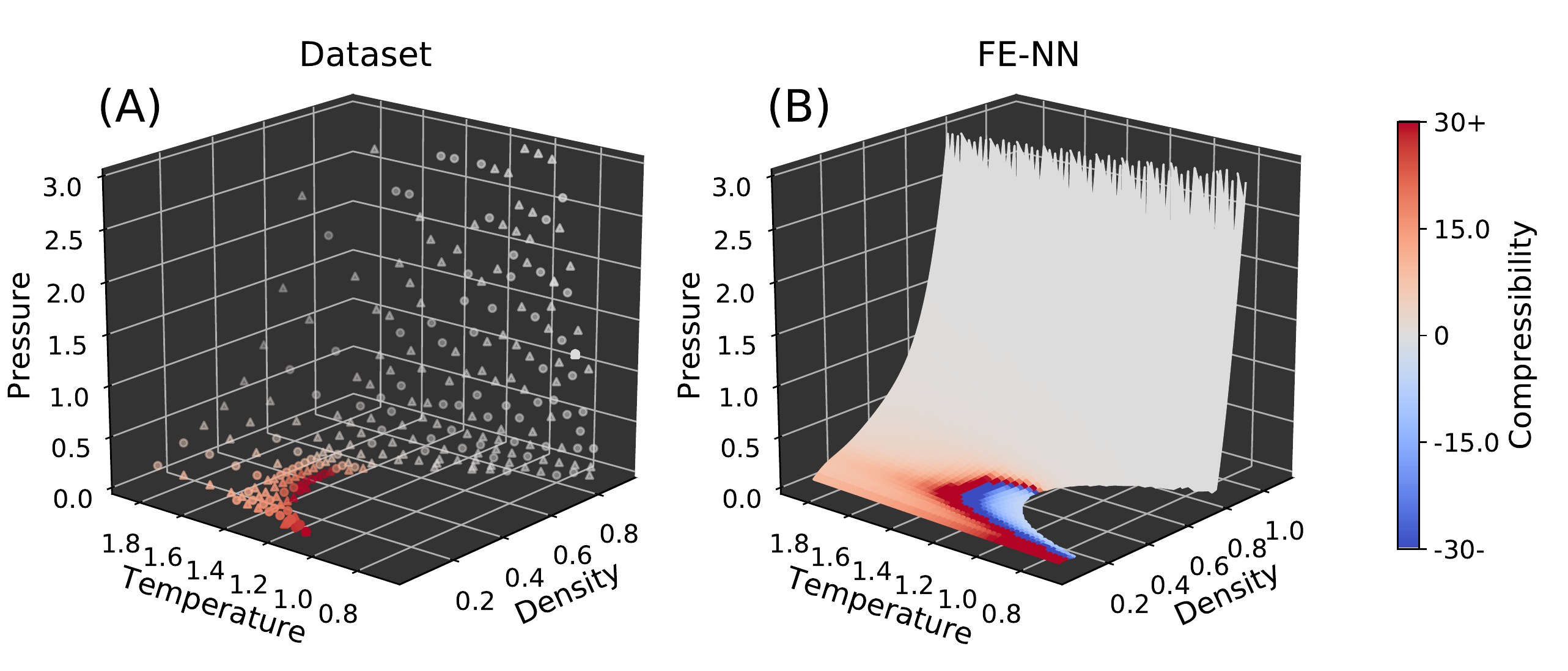}
}
\caption{The LJ system. Color indicates isothermal compressibility $\beta$. (A) Data; Training points (circles) and Validation and Testing points (triangles). (B) FE-NN. Note that the model predicts an unstable region even when no data is available in this region of phase space.}
\label{fig:LJ_phases}
\end{figure*}

\begin{figure} [htbp]
{%
\includegraphics[width=0.8\columnwidth]{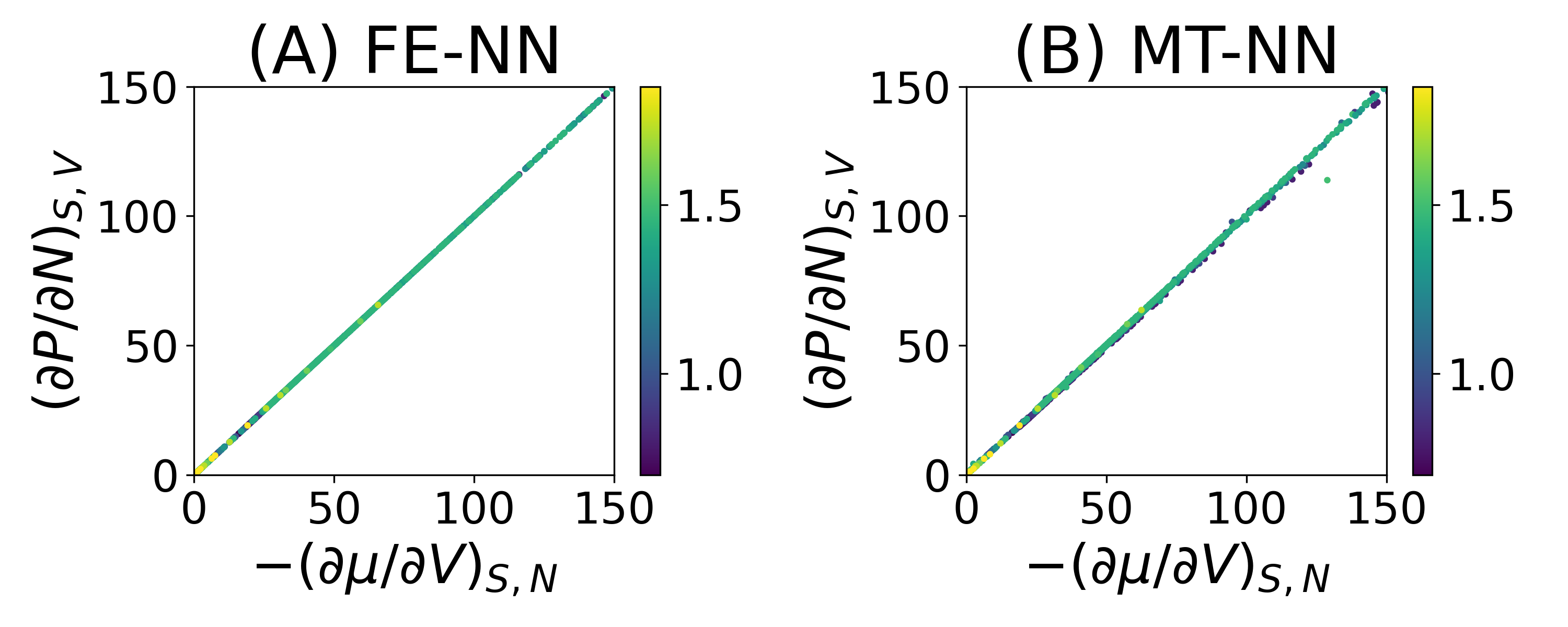}
}
\caption{Maxwell relation for the LJ EOS evaluated from (A) the FE-NN and (B) the MT-NN.}
\label{fig:LJ_Maxwell}
\end{figure}

In order to show the generality of the method, we now move to the more applied example, EOS for Lennard-Jones (LJ) fluids~\cite{thol_equation_2016}.
In Fig.~\ref{fig:LJ_free_ener_cost} (A) we show the loss function of the FE-NN model as a function of the number of epochs. The blue line corresponds to the randomly selected 40\% of the data which has been used for training, whereas the orange line corresponds to left out 30\% of the training data set against which the model is validated. We can see that for both data sets the loss function is well optimized.
In Fig.~\ref{fig:LJ_free_ener_cost} (B) we show the loss function of the MT-NN model as a function of the number of epochs. The MT-NN learner shows acceptable minimization of the loss over the training data. However, for validation it reaches a plateau value, which indicates over-fitting and consequently will lead to poor transferability. We want to emphasize that overfitting does not occur in general for the MT-NN model. We tested different network architectures (see supporting information Tables S~V-VII~\cite{supporting_information}) and different random seeds (see supporting information Table S~VIII~\cite{supporting_information}) to split the data into training, validation and test and only in some cases the overfitting is observed.\newline
In Fig.~\ref{fig:LJ_phases}(A) we illustrate some of the phase space covered. Fig. \ref{fig:LJ_phases}(B) shows the FE-NN reconstruction in this space. 
In Fig.~\ref{fig:LJ_Maxwell} we validate both methods against the Maxwell relation of Eq.~\eqref{Maxwell_rel}, and see again that only the FE-NN can accurately recover it, just as in the VdW system.
Table~\ref{tbl:2} presents the MAE values for both models for first and second order thermodynamic properties on the LJ system. The FE-NN again does consistently better than the MT-NN, especially for the second order derivatives. MAE values are averaged over 4 independent runs using different seed values for train/validation/test splitting (see Table S~VIII in the supporting information~\cite{supporting_information}). The large deviations for $\beta$ and $c_v$ were found to be associated with the critical region (see phase diagrams in Figs.~S4 (A)-(C) and (H)-(I) in the supporting information~\cite{supporting_information}).
This highlights possible grounds for improvement in the model; smooth models can encounter difficulties in the presence of first or second order phase transitions. \newline
\begin{table}
  \caption{Mean absolute error between predicted and true data obtained from the LJ EOS~\cite{stephan_thermophysical_2019}. Thermophysical properties considered are the isothermal compressibility $\beta$, the thermal pressure coefficient $\gamma$, the isochoric heat capacity $c_v$, and the thermal expansion coefficient $\alpha$.}
  \label{tbl:2}
  \begin{tabular}{llllllll}
    \hline
    Model   &$\mu$ & P & U & $\beta$ & $\gamma $  & $c_v$ &$\alpha$  \\
    \hline
    FE-NN & 0.051 & 0.028 & 0.010 & 3.623 & 0.144 & 0.131 & 0.613 \\
    MT-NN  &  0.112 & 0.064 & 0.038 & 14.175 & 0.648 & 0.145  & 7.174    \\
    \hline
  \end{tabular}
\end{table} 

\begin{table}
  \caption{Standard deviations ($\sigma$) for the mean absolute error values reported in Table~\ref{tbl:2}.}
  \label{tbl:S8b}
  \begin{tabular}{llllllll}
    \hline
    Model   & $\sigma_\mu$ & $\sigma_P$   & $\sigma_U$  & $\sigma_\beta$  & $\sigma_\gamma$  & $\sigma_{c_v}$ &$\sigma_\alpha$\\
    \hline
    FE-NN&  1.74E-02 & 1.78E-02 & 5.6E-03 & 1.4631 &7.22E-02 & 6.40E-02 & 2.932-01  \\
    MT-NN  &  6.08E-02 & 3.26E-02 & 2.69E-02 & 8.8227 &  4.455E-01 & 7.00E-02 &3.5797\\
    \hline
  \end{tabular}
\end{table}

\begin{figure*} [htbp]
{%
\includegraphics[width=0.8\textwidth]{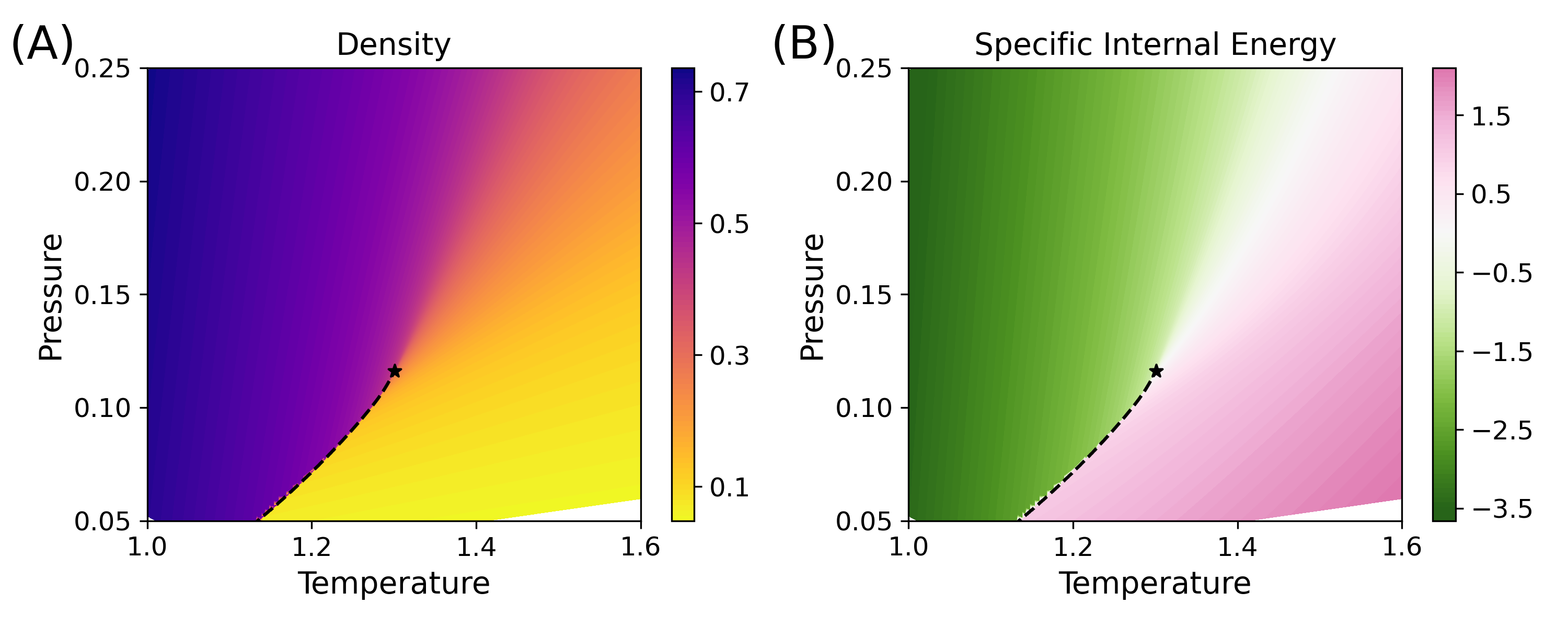}
}
\caption{Results of a FE-NN for LJ data as a function of pressure and temperature. (A): Density ($N/V$), (B): Specific internal energy ($U/N$).  The dashed line indicates where the Maxwell construction was required, corresponding to the first-order phase transition. The star shows the highest temperature value for the dashed line, corresponding to the model's critical point. }
\label{fig:learned_phases}
\end{figure*}
Finally, we present a practical demonstration of the model in Fig.~\ref{fig:learned_phases}. The FE-NN was applied to a $1000\times1000$ grid of points covering the full range of temperatures and densities present in the LJ dataset. This process is extremely fast, taking less than ten seconds on standard laptop CPU. We then used the Maxwell construction numerically to define the pressure as a function of density through the co-existence region for each isotherm. We plot the density and specific internal energy with respect to the pressure and temperature in the vicinity of the critical point. The locations where the Maxwell construction was applied are plotted as a dashed line ending in a star; we emphasize that this line was constructed using the model itself. Immediately visible in the plot is the classic structure of the first order liquid-gas phase transition, and its associated critical point. This recapitulates the classic textbook picture of a low-density gas comprised of unbound particles, a high-density liquid comprised of bound particles, and a continuous super-critical region that connects them. The plots are also smooth, showing that the outputs of the neural network are well-behaved and do not contain any significant high-frequency artifacts.
\section{Conclusion}
In conclusion, we compared a standard Multi-Task model (MT-NN) to a Free Energy Neural Network (FE-NN) trained to predict free energy implicitly using automatic differentiation. We showed that only the Free Energy Neural Network model is able to yield a consistent thermodynamic model that obeys the Maxwell relations. Further, it more accurately learns all thermodynamic properties for both test cases, the VdW and the LJ datasets. Additionally, for the VdW EOS we indeed provided evidence that the FE-NN implicitly learns the free energy and entropy accurately without including them explicitly in the training of the model. Because the Maxwell relations are satisfied, the learned free energy model is suitable for the computation of thermodynamic cycles or the embedding of the model in a larger simulation context (e.g. hydrodynamic or kinetic evolution). This implicit learning of free energies may as well be useful beyond learning EOS. The free energy is a fundamental property in biophysical systems~\cite{gruebele_protein_2002} and for the development of coarse grained models~\cite{noid_perspective:_2013,rice_coarse-graining_2016}. However, to be applicable for those more complex problems further improvements may be desired. To improve the behavior near a phase transition, it is possible to add another layer of automatic differentiation during training in order to include second order partial derivatives in the model construction. Or if the analytic structure of the phase transition is known, future work may allow one to build this structure into the FE-NN, by designing a singularity in the activation functions that corresponds to the universality class of the transition. Generalizing the FE-NN technique to different thermodynamic potentials and to multicomponent systems is straightforward, and the analytical advantages will undoubtedly transfer, but the empirical error of such models has yet to be established. Whereas all these approaches suggest further improvements, the presented FE-NN model is the first step to elegantly learn thermodynamic properties and their generating free energy, without requiring the simulations to measure the free energy itself, and in principle without the need of advanced/enhanced simulation techniques~\cite{van_gunsteren_thermodynamic_1987,wang_efficient_2001,abrams_enhanced_2013,noe_boltzmann_2019}; simulations to measure thermodynamic responses (free energy derivatives) are often more straightforward, can be performed in the most convenient ensemble, and are easily parallelized across the thermodynamic phase space.

\begin{acknowledgments}
The authors thank Mohamed Mehana, David Sanchez, and Galen Craven for fruitful discussions.
This work was supported by the Laboratory Directed Research and Development program of Los Alamos National Laboratory (LANL) under project number 20190005DR. LANL is operated by Triad National Security, LLC, for the National Nuclear Security Administration of U.S. Department of Energy (Contract No. 89233218CNA000001).
DR acknowledges partly funding from the Center for Nonlinear Studies (CNLS) at LANL. KB acknowledges support from the Center of Materials Theory as a part of the Computational Materials  Science  (CMS)  program,  funded  by  the  U.S. Department of Energy, Office of Basic Energy Sciences.
\end{acknowledgments}

\bibliography{literature}
\end{document}


\preprint{}

\title{Supporting information: Machine Learning of consistent thermodynamic models using automatic differentiation.}

\affiliation{Los Alamos National Laboratory, Theoretical Division, Chemistry and Physics of Materials Group, Los Alamos, 87545 New Mexico (USA). }
\affiliation{ 
Los Alamos National Laboratory, Computer, Computational \& Stat Sciences Division,  Information Sciences Group, Los Alamos, 87545 New Mexico (USA).}
\author{David Rosenberger}
 \email{d.rosenberger@fu-berlin.de}
 \altaffiliation[Current affiliation:~]{Freie Universit\"at Berlin (Germany).}
 \affiliation{Los Alamos National Laboratory, Theoretical Division, Chemistry and Physics of Materials Group, Los Alamos, 87545 New Mexico (USA). }
\author{Nicholas Lubbers}%
\email{nlubbers@lanl.gov}
\affiliation{ 
Los Alamos National Laboratory, Computer, Computational \& Stat Sciences Division,  Information Sciences Group, Los Alamos, 87545 New Mexico (USA).}
\author{Kipton Barros}
\affiliation{Los Alamos National Laboratory, Theoretical Division, Chemistry and Physics of Materials Group, Los Alamos, 87545 New Mexico (USA). }
\author{Timothy C. Germann}
\affiliation{Los Alamos National Laboratory, Theoretical Division, Chemistry and Physics of Materials Group, Los Alamos, 87545 New Mexico (USA). }

\maketitle


Data generation and data analysis is based on open source python packages. We utilized the numpy library for basic numerical operations~\cite{harris2020array}.
Training of the neural networks has been performed with pytorch~\cite{paszke_pytorch_2019}.
The data for the Lennard-Jones systems has been extracted using the openpyxl library~\cite{gazoni_eric_openpyxl_nodate} in combination with pandas dataframes~\cite{mckinney-proc-scipy-2010,reback2020pandas}.
The corresponding figures have been generated with matplotlib~\cite{Hunter:2007}.

\newpage
\section{van der Waals Equation of state}

In order to assess the robustness of the machine learning results, we tested different machine learning architectures, i.e. different number of hidden layers and different number of neurons per hidden layer. In Tables S~\ref{tbl:S1}-S~\ref{tbl:S3} we present the mean absolute errors for training, validation and test.
\begin{table}[hb]
  \caption{Mean absolute error between predicted and true training data generated from the van der Waals EOS using different network architectures. Thermophysical properties considered are, the chemical potential $\mu$, the pressure $P$, the internal energy $U$, entropy $S$, and free energy $A$.}
  \label{tbl:S1}
  \begin{tabular}{lllllllll}
    \hline
    Model  & hidden layers & neurons/layer & $\mu$ & $P $  & $U$  &$S$ &$A$ \\
    \hline
    FE-NN& 1 & 8 & 0.004 & 0.001 & 0.003 & 0.011 & 0.002\\
    MT-NN  & 1 & 8 & 0.016 & 0.002 & 0.001 & &\\
    FE-NN& 2 & 8 & 0.001 & 0.0 & 0.001 & 0.003 & 0.0\\
    MT-NN  & 2 & 8 & 0.003 & 0.001 & 0.001 & &\\
    FE-NN& 2 & 16 & 0.001 & 0.0 & 0.001 & 0.002 & 0.0\\
    MT-NN  & 2 & 16 & 0.004 & 0.001 & 0.001 & & \\
    FE-NN& 3 & 16 & 0.0 & 0.0 & 0.0 0.0 & 0.001 & 0.0\\
    MT-NN  & 3 & 16 & 0.004 & 0.001 & 0.0 & &\\
    \hline
  \end{tabular}
\end{table}

\begin{table}
  \caption{Mean absolute error between predicted and true validation data generated from the van der Waals EOS using different network architectures. Thermophysical properties considered are, the chemical potential $\mu$, the pressure $P$, the internal energy $U$, entropy $S$, and free energy $A$..}
  \label{tbl:S2}
  \begin{tabular}{lllllllll}
    \hline
    Model  &hidden layers & neurons/layer & $\mu$ & $P $  & $U$  &$S$ &$A$ \\
    \hline
    FE-NN& 1 & 8 & 0.004 & 0.001 & 0.002 &  0.006 & 0.002\\
    MT-NN  & 1 & 8 & 0.013 & 0.001 & 0.001 & &\\
    FE-NN& 2 & 8 & 0.001 & 0.0 & 0.001 & 0.003 & 0.0\\
    MT-NN  & 2 & 8 & 0.003 & 0.001 & 0.0013 & &\\
    FE-NN& 2 & 16 & 0.0 & 0.0 & 0.001 & 0.001 & 0.0\\
    MT-NN  & 2 & 16 & 0.005 & 0.001 & 0.0 & &\\
    FE-NN& 3 & 16 & 0.0 & 0.0 & 0.001 & 0.003 & 0.0\\
    MT-NN  & 3 & 16 & 0.004 & 0.0 & 0.0 & & \\
    \hline
  \end{tabular}
\end{table}

\begin{table}
  \caption{Mean absolute error between predicted and true test data generated from the van der Waals EOS using different network architectures. Thermophysical properties considered are, the chemical potential $\mu$, the pressure $P$, the internal energy $U$, entropy $S$, and free energy $A$.}
  \label{tbl:S3}
  \begin{tabular}{lllllllll}
    \hline
    Model  & hidden layers & neurons/layer & $\mu$ & $P $  & $U$  &$S$ &$A$ \\
    \hline
    FE-NN& 1 & 8 & 0.005 & 0.001 &  0.003 & 0.011 & 0.002\\
    MT-NN  & 1 & 8 &  0.015 & 0.002 & 0.001 & &\\
    FE-NN& 2 & 8 & 0.001 & 0.0 & 0.002 & 0.004 & 0.0\\
    MT-NN  & 2 & 8 &  0.004 & & 0.001 & 0.001 & &\\
    FE-NN& 2 & 16 &  0.001 & 0.0 & 0.001 & 0.003 & 0.0\\
    MT-NN  & 2 & 16 & 0.005 & 0.0 & 0.0 & & \\
    FE-NN& 3 & 16 & 0.001 & 0.0 & 0.0 & 0.001 & 0.0\\
    MT-NN  & 3 & 16 & 0.005 & 0.001 & 0.0 & & \\
    \hline
  \end{tabular}
\end{table}

From the MAE values one can see that in principle all architectures produce very low MAE values. However, in combination with the Lennard-Jones data set (see Tables S~\ref{tbl:S5}- S~\ref{tbl:S7}), we choose to take 3 hidden layers with 16 neurons in order to use the same architecture for both problems. \newline
Besides different architectures, we also tested how stable the results are by using different seed numbers to split the complete data set into training, validation and test. In Table S~\ref{tbl:S4}, we show the corresponding mean absolute error values also for the kernel ridge regression (KRR) model and see that the findings presented in the main article are robust against which sub-set of the data is used for training.
\begin{table}
  \caption{Mean absolute error between predicted and true data generated from the van der Waals EOS using different seed numbers for splitting the data. Thermophysical properties considered are, the chemical potential $\mu$, the pressure $P$, the internal energy $U$,entropy $S$, free energy $A$, and the isothermal compressibility $\beta$.}
  \label{tbl:S4}
  \begin{tabular}{llllllll}
    \hline
    Random Seed &Model   & $P$ & $\mu$   & $U$ & $A$ & $S$ & $\beta$  \\
    \hline
    70201987 & FE-NN &  0.0002 & 0.0005 & 0.0005 & 0.0002 & 0.0019 & 0.1720  \\
    & MT-NN  &  0.0008 & 0.0044 & 0.0006 & &  & 0.5702 \\
    & KRR & 0.0077 & 0.0046 & 0.0002 & & &14.2620\\
    \hline
    42 &FE-NN y&  0.0003 & 0.0013  & 0.0009 & 0.0004 & 0.0020 & 0.3564  \\
    & MT-NN  &  0.0020 & 0.0169 & 0.0025 & &  &0.6159  \\
    & KRR & 0.0072 & 0.0037 & 0.0002 & & & 11.9177\\
    \hline
     1337 &FE-NN &  0.0003 & 0.0010 & 0.0015 & 0.0010 & 0.0059 &  0.1976 \\
    & MT-NN  &  0.0060 &  0.0035 & 0.0120 & &  & 2.6610 \\
    & KRR & 0.0088 & 0.0035 &  0.0002 & & & 24.6415\\
    \hline
    801945  & FE-NN &  0.0002 & 0.0008 & 0.0010 & 0.0002  & 0.0029 & 0.2065 \\
    &MT-NN &  0.0004 & 0.0042  & 0.0015  & &  & 0.2607  \\
     & KRR &  0.0072 & 0.0035 & 0.0002 & & &14.3442\\
    \hline
  \end{tabular}
\end{table}

\begin{figure} [htbp]
{%
\includegraphics[width=0.9\columnwidth]{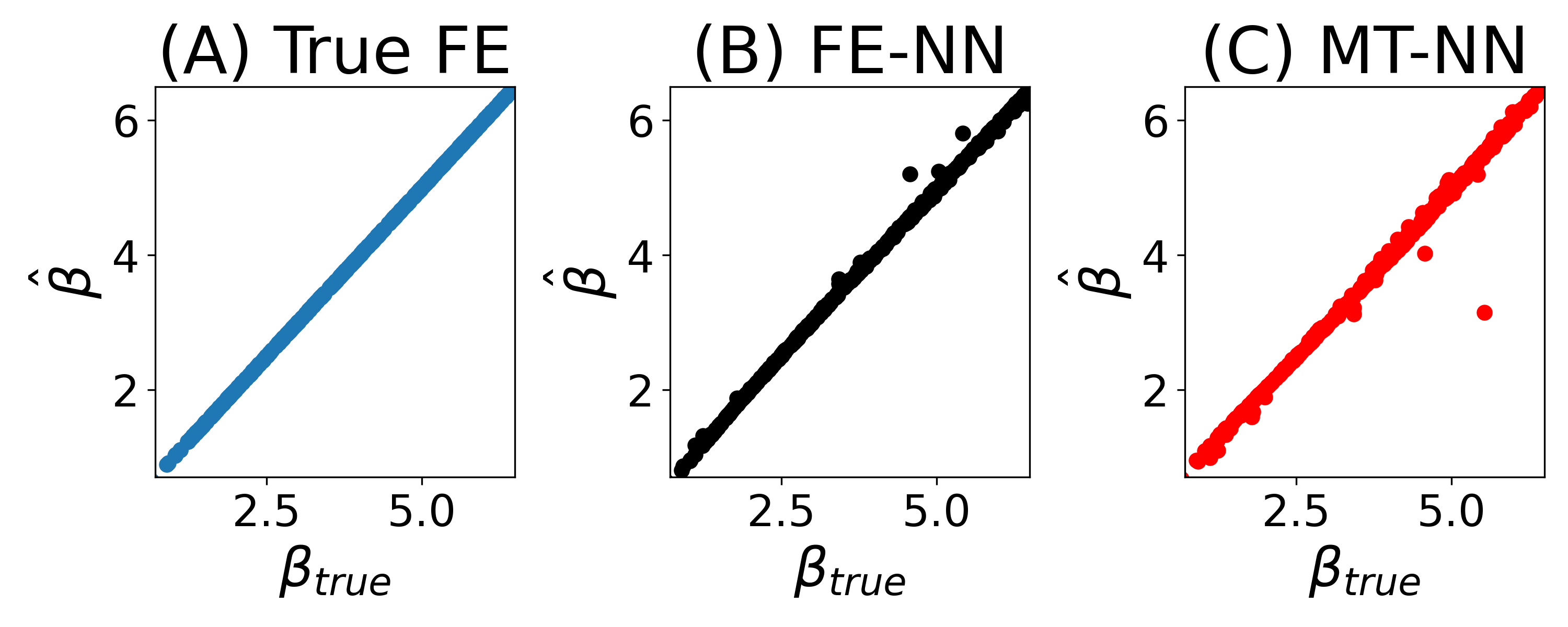}
}
\caption{Correlation between the analytical isothermal compressibility for the van der Waals EOS and the isothermal compressibility obtained from AD of the pressure with respect to volume, where pressure has been obtained from (A) the true free energy, (B) the FE-NN and (C) the MT-NN.}
\label{fig:vdW_beta}
\end{figure}

To validate the automatic differentiation (AD) and to test its application for second order derivatives we assess the performance of the two neural network based models on the isothermal compressibility ${\beta}$.
In Figure S\ref{fig:vdW_beta}(A) $\hat{\beta}$ is obtained from AD of the pressure derived from the true free energy. As expected, there is no difference between the predicted and true isothermal compressibility, validating the AD procedure. In Fig.~S\ref{fig:vdW_beta}(B) one sees that results for the FE-NN are more accurate than for the MT-NN (see Fig.~S\ref{fig:vdW_beta}(C)), which is quantitatively indicated from the MAE of $\beta$ (see Table I in the main article), which is about 5 times larger for the MT-NN compared to the FE-NN.
\newpage
\section{Lennard-Jones Equation of state}

In Tables S~\ref{tbl:S5}-S~\ref{tbl:S7}, we present the MAE values for training, validation and testing for different architectures. The MT-NN model is showing slightly better training and validation performance compared to the FE-NN model especially for the smaller network architectures. The performance for the FE-NN model significantly improves if the ANN is 2 hidden layers deep and 16 neurons wide. If we add one hidden layer the performance improves further. This leads to a significant better performance of the FE-NN model on the test set. (see Table S~\ref{tbl:S7}).

\begin{table}[h]
  \caption{Mean absolute error between predicted and true training data generated from the Lennard-Jones EOS using different network architectures. Thermophysical properties considered are, the chemical potential $\mu$, pressure $P$ and the internal energy $U$.}
  \label{tbl:S5}
  \begin{tabular}{lllllll}
    \hline
    Model  & hidden layers & neurons/layer & $\mu$ & $P $  & $U$  \\
    \hline
    FE-NN& 1 & 8 &  0.114  & 0.074  & 0.058\\
    MT-NN  & 1 & 8 &  0.038  & 0.054 & 0.009\\
    FE-NN& 2 & 8 &  0.118  & 0.065  & 0.038\\
    MT-NN  & 2 & 8 &  0.112 & 0.032& 0.013\\
    FE-NN& 2 & 16 & 0.064 & 0.068 & 0.049\\
    MT-NN  & 2 & 16 & 0.043  & 0.046 & 0.014  \\
    FE-NN& 3 & 16 & 0.007 & 0.005 & 0.002\\
    MT-NN  & 3 & 16 & 0.014 & 0.013 &  0.008 \\
    \hline
  \end{tabular}
\end{table}

\begin{table}
  \caption{Mean absolute error between predicted and true validation data generated from the Lennard-Jones EOS using different network architectures. Thermophysical properties considered are, the chemical potential $\mu$, pressure $P$ and the internal energy $U$.}
  \label{tbl:S6}
  \begin{tabular}{lllllll}
    \hline
    Model  & hidden layers & neurons/layer & $\mu$ & $P $  & $U$  \\
    \hline
    FE-NN& 1 & 8 &   0.077 & 0.065  & 0.052 \\
    MT-NN  & 1 & 8 &   0.049 &  0.069 & 0.011  \\
    FE-NN& 2 & 8 &   0.082 &  0.046 & 0.037 \\
    MT-NN  & 2 & 8 &   0.044  & 0.035 & 0.014 \\
    FE-NN& 2 & 16 &  0.026  & 0.05  & 0.051 \\
    MT-NN  & 2 & 16 &  0.045  & 0.049 & 0.016 \\
    FE-NN& 3 & 16 &  0.075 & 0.021 & 0.006\\
    MT-NN  & 3 & 16 & 0.092 & 0.027 & 0.009\\
    \hline
  \end{tabular}
\end{table}
\begin{table}
  \caption{Mean absolute error between predicted and true test data generated from the Lennard-Jones EOS using different network architectures. Thermophysical properties considered are, the chemical potential $\mu$, pressure $P$ and the internal energy $U$.}
  \label{tbl:S7}
  \begin{tabular}{lllllll}
    \hline
    Model  & hidden layers & neurons/layer & $\mu$ & $P $  & $U$  \\
    \hline
    FE-NN& 1 & 8 & 0.079 & 0.073 &  0.057\\
    MT-NN  & 1 & 8 &  0.112 & 0.063 & 0.01 \\
    FE-NN& 2 & 8 & 0.085 & 0.061 &  0.038\\
    MT-NN  & 2 & 8 &   0.112 & 0.032 & 0.013\\
    FE-NN& 2 & 16 & 0.031 & 0.06  & 0.05\\
    MT-NN  & 2 & 16 & 0.112 & 0.05  & 0.015\\
    FE-NN& 3 & 16 &  0.007 & 0.005 & 0.003 \\
    MT-NN  & 3 & 16 & 0.02 & 0.016 & 0.006  \\
    \hline
  \end{tabular}
\end{table}
Besides the different architectures, we also tested different seed values to generate the random split of the data into training, validation and test (see Table S~\ref{tbl:S8}).
\begin{table}
  \caption{Mean absolute error between predicted and true data generated from the Lennard-Jones EOS using different seed numbers for splitting the data. Thermophysical properties considered are, the chemical potential $\mu$, pressure $P$, internal energy $U$, isothermal compressibility $\beta$, the thermal pressure coefficient $\gamma$, the isochoric heat capacity $c_v$. and the thermal expansion coefficient $\alpha$.}
  \label{tbl:S8}
  \begin{tabular}{lllllllll}
    \hline
    Random Seed & Model   &$\mu$ & $P$ & $U$ & $\beta$ & $\gamma $  & $c_v$ &$\alpha$  \\
    \hline
     70201987 &FE-NN & 0.0630 & 0.0469 & 0.0152 & 5.7737 & 0.2186  & 0.0713  & 0.1902\\
             &MT-NN & 0.0980 & 0.0927 & 0.0696 &  9.0008 & 1.16923 & 0.2170 & 8.0745 \\
    \hline
    42 &FE-NN &    0.066  & 0.0394 & 0.0138  & 2.501 & 0.180 & 0.149 & 0.646  \\
       &MT-NN & 0.182 & 0.0714 & 0.0266 & 5.194 & 0.447 & 0.092 & 1.519 \\
    \hline
    1337 &FE-NN & 0.048 & 0.016 & 0.008 & 3.179  &  0.213 & 0.126 & 0.780  \\
         &MT-NN & 0.068 & 0.043 & 0.022 & 8.739 & 0.406 & 0.146 & 5.740 \\
    \hline
    801945 & FE-NN & 0.028 & 0.010  & 0.003 & 3.040  &  0.052 & 0.090 & 0.837 \\
           & MT-NN & 0.042 & 0.018   & 0.007 & 24.945 & 0.124 & 0.057 & 9.782\\
    \hline
  \end{tabular}
\end{table}

\begin{figure} [htbp]
{%
\includegraphics[scale=0.5]{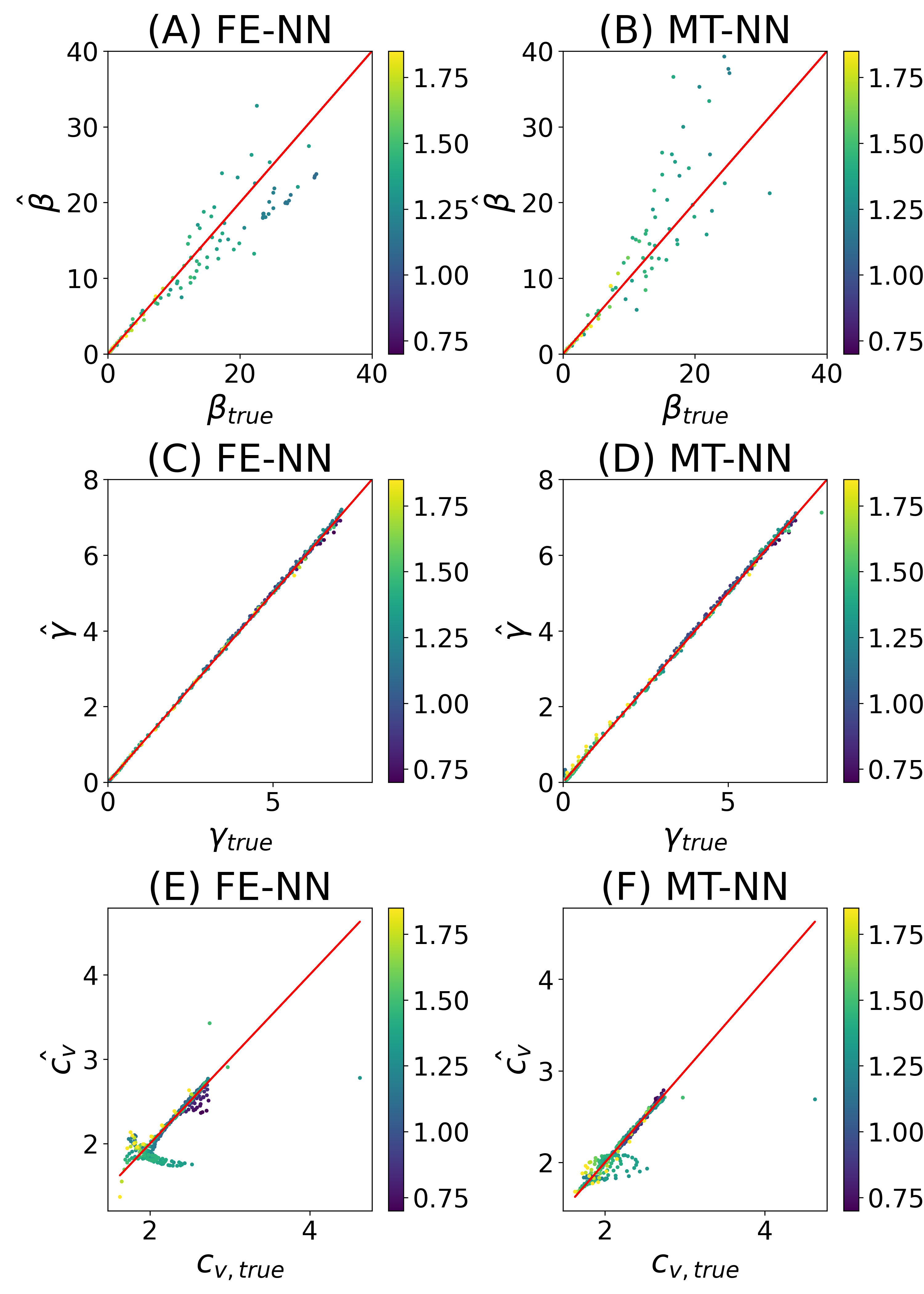}
}
\caption{Left panel: Correlation between the learned properties. (A) Isothermal compressibility $\hat{\beta}$, (C) thermal pressure coefficient $\hat{\gamma}$ and (E) isochoric heat capacity $c_v$ obtained from the FE-NN against the true values. Right panel: corresponding results for the MT-NN model, (B) $\hat{\beta}$, (D) $\hat{\gamma}$ and (F) $c_v$. The red line shows where true and predicted values are equal.}
\label{fig:LJ_correlation}
\end{figure}

\begin{figure} [htbp]
{%
\includegraphics[width=0.6\columnwidth]{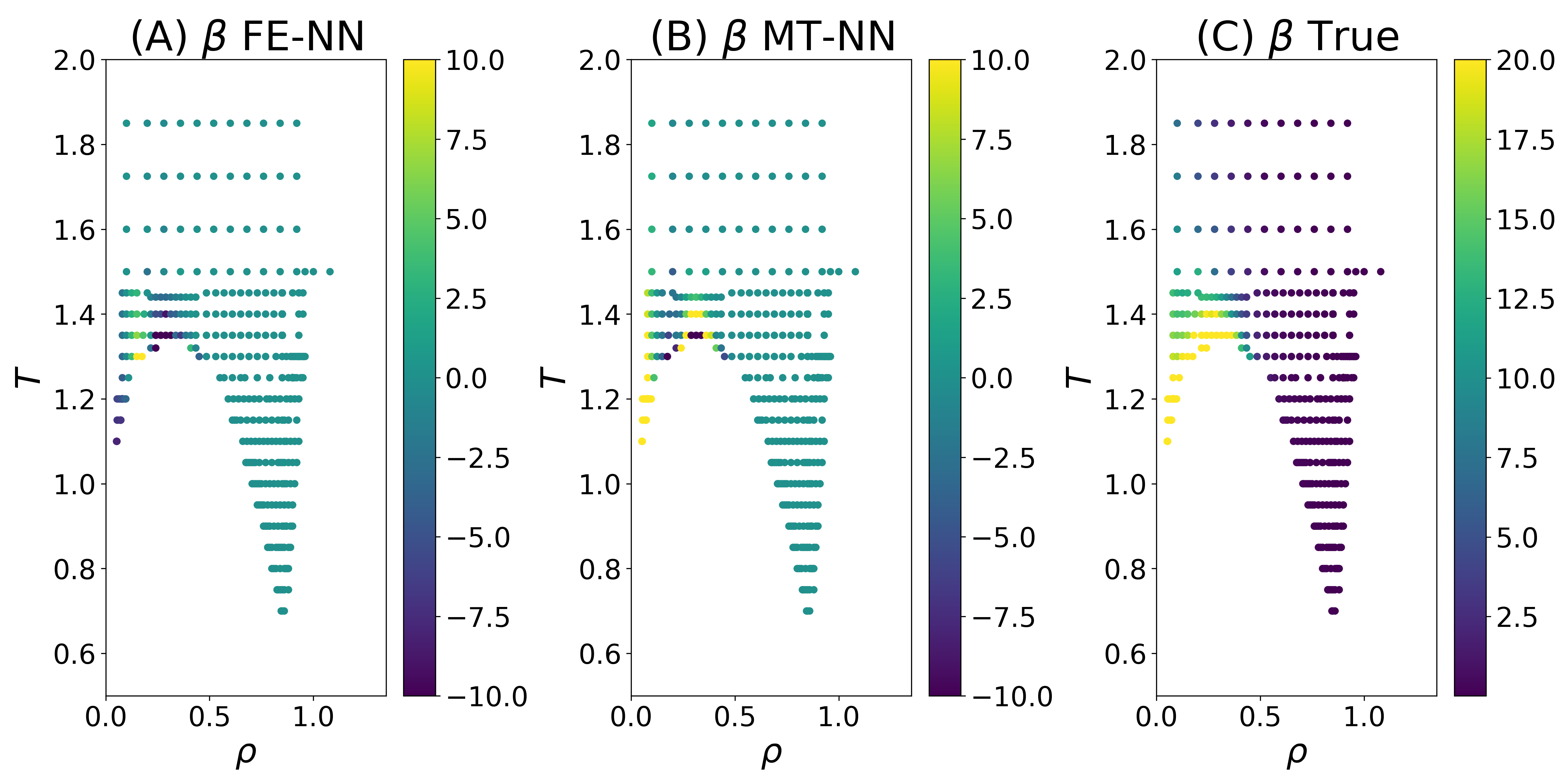}
\includegraphics[width=0.6\columnwidth]{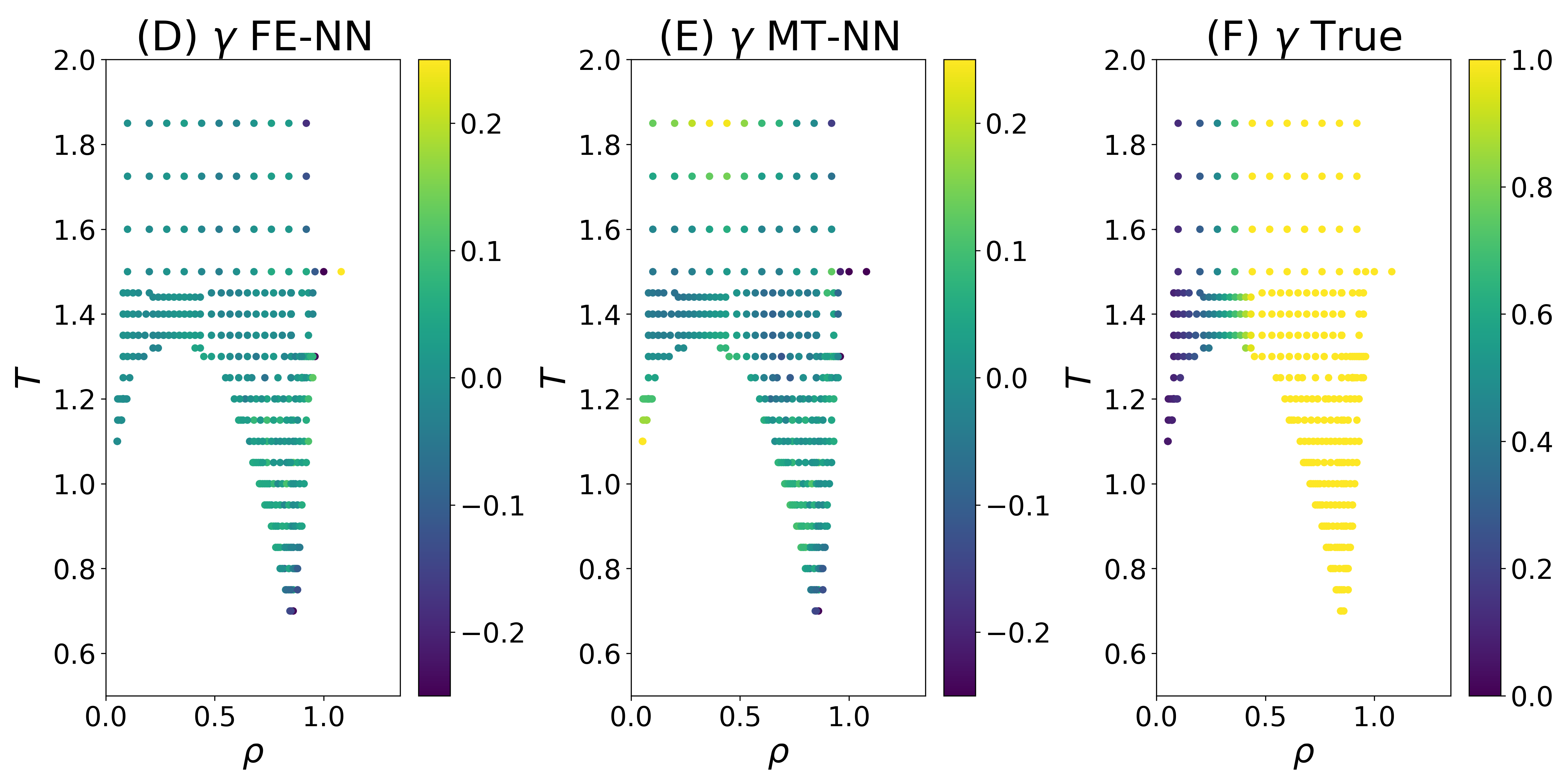}
\includegraphics[width=0.6\columnwidth]{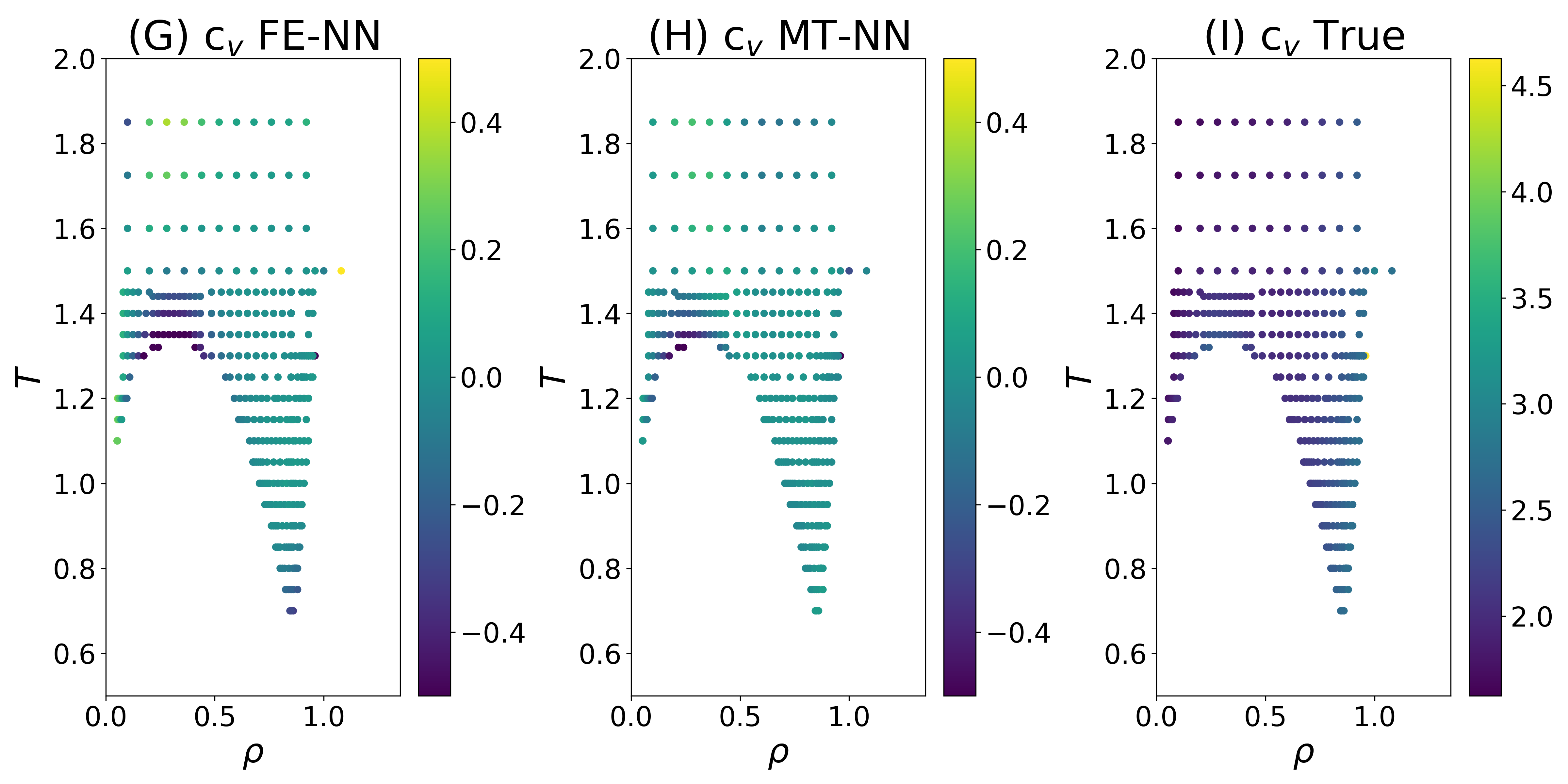}
}
\caption{Phase diagram for isothermal compressibility $\beta$ (upper panel); thermal pressure coefficient $\gamma$ (central panel); isochoric heat capacity $c_V$ (bottom panel). The left panel corresponds to the FE-NN model, the central panel corresponds to the MT-NN model and the right panel depicts the true data obtained from the data set. The color bar indicates the actual value of the thermophysical property under consideration for the true model; for the ML models the color bar represents the difference between the predicted value and the true value.}
\label{fig:LJ_phase_diagrams}
\end{figure}

 Further, in Fig.~S\ref{fig:LJ_correlation}) we show the correlation between the second order thermophysical properties $\beta$, $\gamma$, and $c_v$, and their ground truth. Both models suffer somewhat in the regions of large compressibility $\beta$. The thermal pressure coefficient $\gamma$ is reproduced well by both models, but considerably better by the FE-NN. The heat capacity $c_v$ is predicted well for many state points, but both models suffer in a particular region of the phase space. The deviations for $\beta$ and $c_v$ were found to be associated with the critical region (see phase diagrams in Figs.~S\ref{fig:LJ_phase_diagrams} (A)-(C) and (H)-(I)).\newline

\begin{figure} [htbp]
{%
\includegraphics[width=0.7\columnwidth]{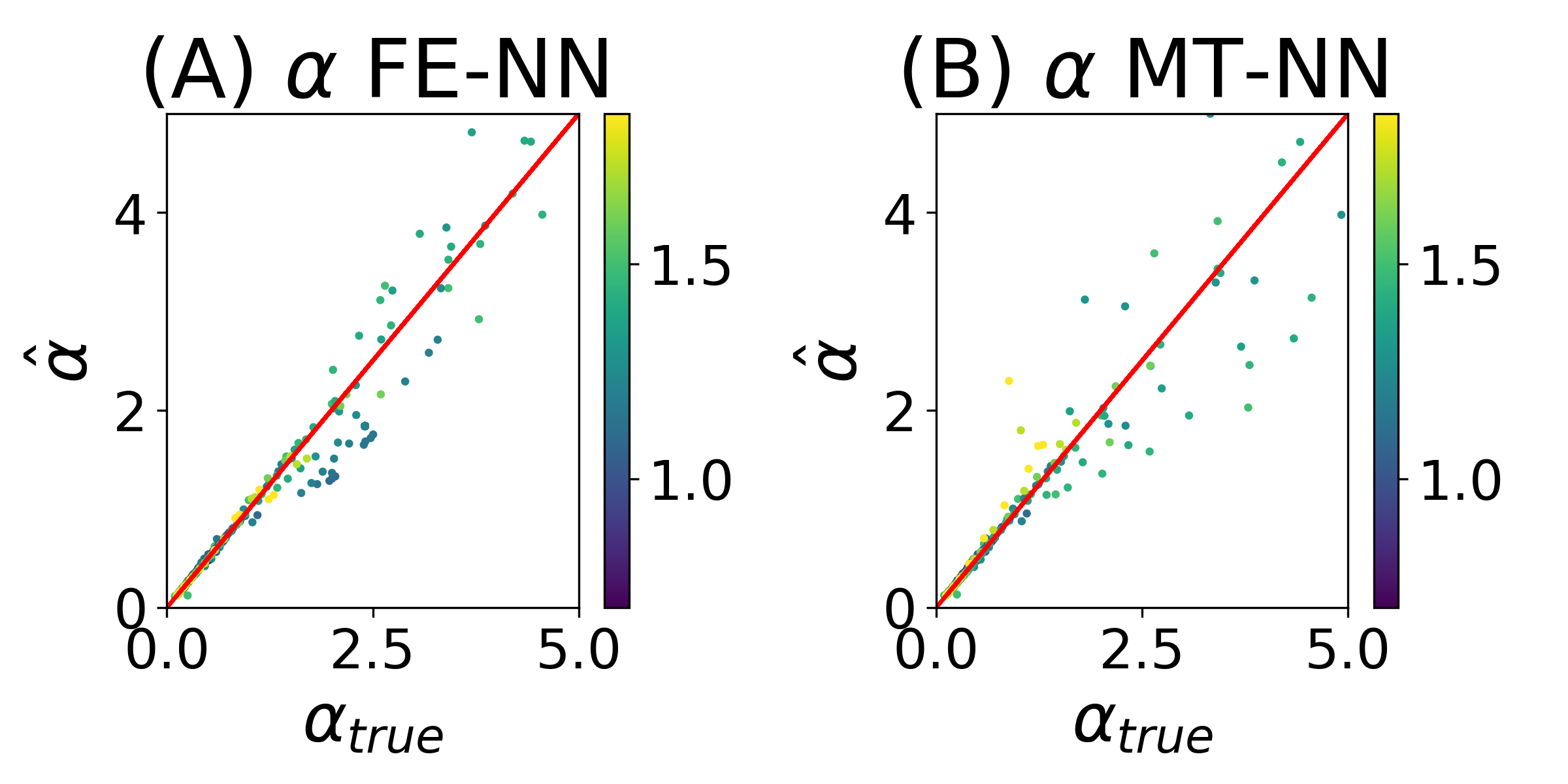}
}
\caption{Correlation between the learned thermal expansion coefficient $\alpha$ and the true value (A): obtained from the FE-NN model; (B) obtained from the MT-NN model. The red line indicates the performance of a perfect model.}
\label{fig:LJ_thermal_expansion}
\end{figure}

For completeness: In Fig. ~S\ref{fig:LJ_thermal_expansion}, we show the correlation plot for the thermal expansion coefficient, which is the product of $\beta$ and $\gamma$. As both models can learn $\gamma$ quite accurately, this plot reflects the mismatch in reproducing $\beta$ as shown in the main article.
\newpage
In Fig.~S\ref{fig:LJ_phase_diagrams} we show the phase diagram, $T$ vs. $\rho$, where for the true model the color bar corresponds to the value of the corresponding thermophysical property and for the FE-NN and MT-NN model the color bar represents the difference between the predicted and the true value. In the upper panel the isothermal compressibility $\beta$ is depicted, in the central panel we present the thermal pressure coefficient $\gamma$ and in the bottom panel we show the isochoric heat capacity $c_V$. The left panel corresponds to the FE-NN model, the central panel corresponds to the MT-NN model and the right panel is the true data obtained from the data set~\cite{stephan_thermophysical_2019}.
From the phase diagrams we see that both ML models show poor predictions for $\beta$ around the critical point. For $\gamma$ both models yield equal predictions and are in good agreement with the true data. For $c_v$ the predictions are again poor around the critical point.\cite{stephan_thermophysical_2019}

\newpage
\bibliography{literature}